\documentclass[showpacs,preprintnumbers,superscriptaddress,amsmath,amssymb]{revtex4-2}
\usepackage[colorlinks=true,bookmarks=false,citecolor=blue,urlcolor=blue]{hyperref}
\usepackage{amsmath,amssymb}
\usepackage{graphicx}
\usepackage{verbatim}
\usepackage{enumitem}
\usepackage{color}
\usepackage[english]{babel}
\usepackage{bm}
\usepackage{natbib}
\usepackage[symbol]{footmisc}
\usepackage{textcomp}

\usepackage{epstopdf}
\usepackage{setspace}
\begin{document}
\title{Exceptional points in dielectric spheroid}

\author{Evgeny Bulgakov}
\affiliation{Kirensky Institute of Physics Federal Research Center
KSC SB RAS 660036 Krasnoyarsk Russia} \affiliation{Reshetnev
Siberian State University of Science and Technology, 660037,
Krasnoyarsk, Russia}
\author{Konstantin Pichugin}
\affiliation{Kirensky Institute of Physics Federal Research Center
KSC SB RAS 660036 Krasnoyarsk Russia}
\author{Almas Sadreev}
\affiliation{Kirensky Institute of Physics Federal Research Center
KSC SB RAS 660036 Krasnoyarsk Russia}
\date{\today}

\begin{abstract}
Evolution of resonant frequencies and resonant modes as dependent
on the aspect ratio is considered in a dielectric high index
spheroid. Because of rotational symmetry of the spheroid the
solutions are separated by the azimuthal index $m$. By the
two-fold variation of a refractive index and the aspect ratio we
achieve exceptional points (EPs) at which the resonant frequencies
and resonant modes are coalesced in the sectors $m=0$ for both TE and TM polarizations and $m=1$.
\end{abstract}
 \maketitle

\section{Introduction}
Optical properties of dielectric particle is described by resonant
frequencies and corresponding resonant modes. The most famous case
is a dielectric sphere whose resonant modes and frequencies were
first considered by Stratton \cite{Stratton}. The solutions in the
form quasi-normal modes  (QNMs) leaking from the sphere were
considered in Refs. \cite{Conwell1984,Lai1990a}.
The frequencies of these solutions resonances are complex because
of coupling of the dielectric particle with the radiation
continuum and can be considered as the eigenvalues of the
non-Hermitian Hamiltonian
\cite{Ching1998,Leung1998,Okolowicz2003,Lalanne2018}. Material
losses as well as thermal fluctuations \cite{Lai1990} of
dielectric particle can considerably contribute into the imaginary
part of complex resonant frequencies through complex refractive
index. Non-Hermitian phenomena  drastically alters the behavior of
a system compared to its Hermitian counterpart describing the
closed system. The best example of such a difference is the
avoided resonant crossing (ARC) because of coupling of a particle
with the radiation continuum
\cite{Heiss2000,Rotter2005,Wiersig2006,Bernier2018,Park2018}. In
turn the ARC can emerge to singularities, bound states in the
continuum at which the imaginary part of resonance turns to zero
\cite{Friedrich1985,Sadreev2021} that gives rise to collapse of
Fano resonance and exceptional points (EPs). The last is
remarkable by that  complex frequencies become degenerate and the
eigenmodes coalesce \cite{Heiss1990,Heiss1999,Eleuch2013}. Early
experiments on microwave coupled resonators  revealed the peculiar
topology of eigenvalue surfaces near exceptional points for
encircling of EP \cite{Dembowski2001}.

Many works on EPs and their applications are associated with
parity-time (PT) symmetric optical systems with a balanced gain
and loss. In that case, EPs can be easily found by tuning a single
parameter, namely, the amplitude of the balanced gain and loss
\cite{Brandstetter2014,Longhi2017,Feng2017,Oezdemir2019,Miri2019}.
Since it is not always easy or desirable to keep a balanced gain
and loss in an optical system there is of significant interest to
explore EPs and their applications in non-PT-symmetric optical
systems.  Currently, there exist studies concerning EPs for
resonant states in extended periodic dielectric structures
sandwiched between two homogeneous half-spaces
\cite{Zhen2015,Kaminski2017,Abdrabou2018,Abdrabou2020}, dual-mode
planar optical waveguides \cite{Ghosh2016} and plasmonic waveguide
\cite{Min2020}, layered structures
\cite{Feng2013,Gomis2019,Popov2019}, two infinitely long
dielectric cylinders
\cite{Ryu2012,Kullig2018,Yi2019,Huang2019a,Abdrabou2019} and even
single rod with deformed cross-section
\cite{Unter2008,Kullig2016,Kullig2018a,Yi2019,Jiang2019}. As for
compact dielectric resonators we distinguish the only study of
EPs in compact coated dielectric sphere \cite{Jiang2020}.

In the present paper we consider similar compact elementary
dielectric resonator such as a spheroid in which EPs can be
achieved by two-fold variation of aspect ratio and refractive
index. Although the spheroid allows the solution due to separation
of variables in spheroidal coordinate system
\cite{Asano1975,Barber1975}, analytical expressions for
solutions are too cumbersome. We use software package COMSOL
Multiphysics which allows to obtain numerically the complex
resonant frequencies and corresponding resonant modes of particle
of arbitrary shape embedded into the radiation continuum by use of
perfectly absorbing boundary conditions.

\section{Evolution of resonant frequencies in spheroid}

A rotational symmetry of spheroid implies that the azimuthal index
$m$ is preserved. That allows to calculate the resonant
frequencies and resonant eigenmodes separately in each sector $m$
and calculate EM field configurations as series over the orbital
momenta outside spheroid \cite{Asano1975}
\begin{equation}\label{series}
    \overrightarrow{E}^{(m)}(\overrightarrow{r})=\sum_{l=1}^{\infty}[a_{lm}\overrightarrow{M}_{lm}
(\overrightarrow{r})+b_{lm}\overrightarrow{N}_{lm}(\overrightarrow{r})]
\end{equation}
where $\overrightarrow{M}_{lm}$ and $\overrightarrow{N}_{lm}$ are
the spherical harmonics \cite{Stratton}. In what follows we
consider the sectors $m=0$ and $m=1$.

\begin{figure}
\includegraphics*[width=8cm,clip=]{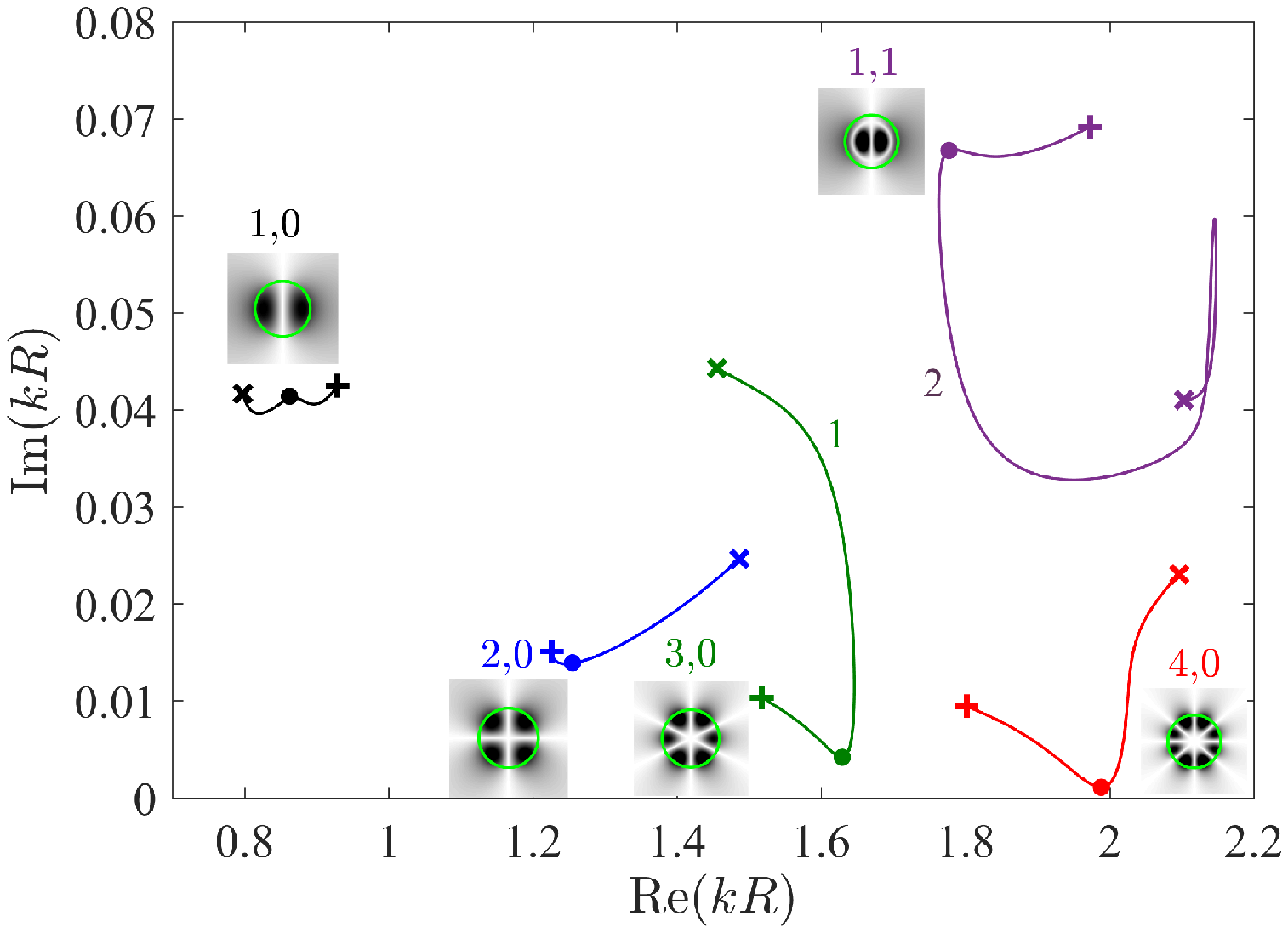}
\includegraphics*[width=8cm,clip=]{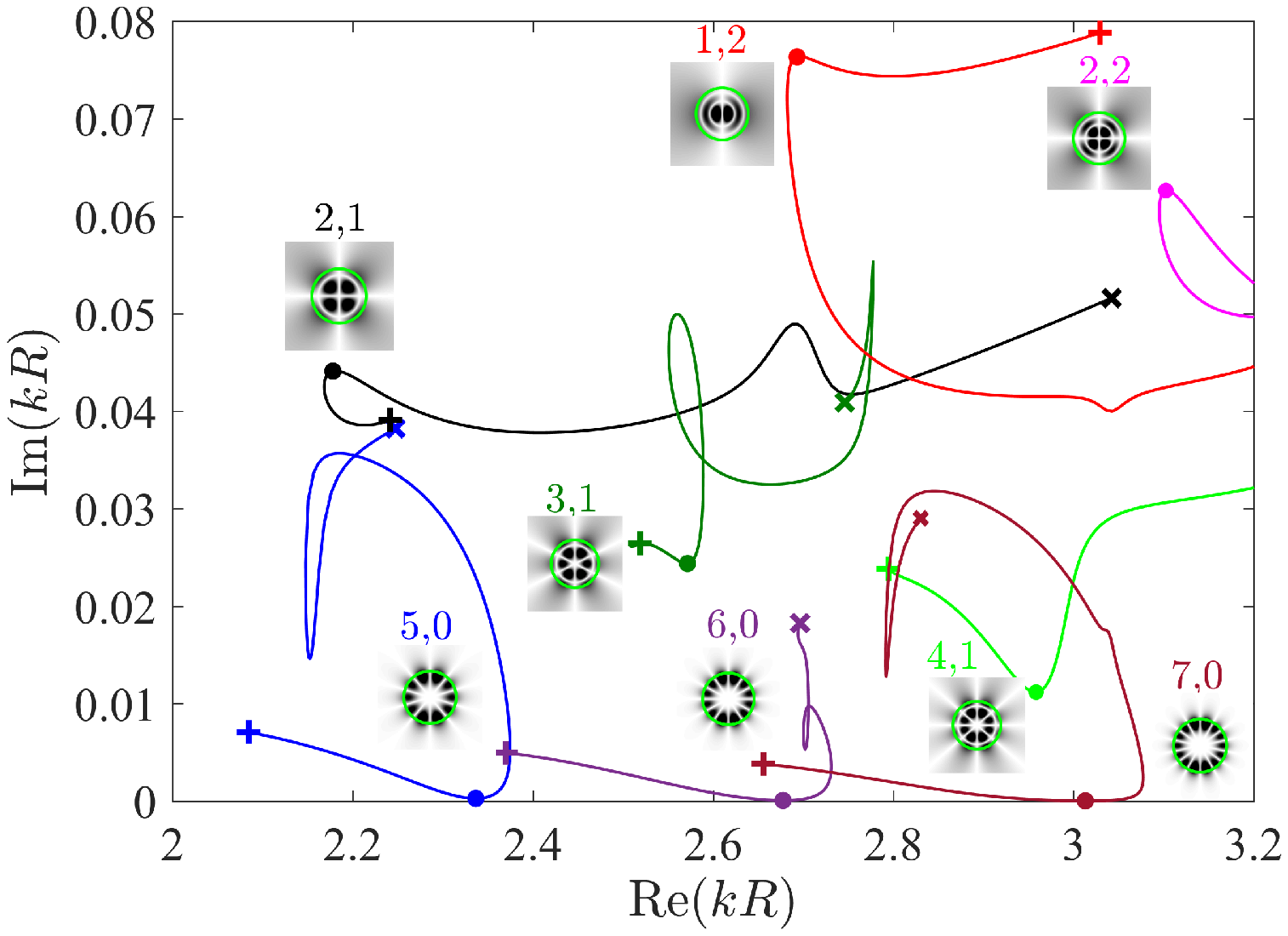}
\caption{Evolution of complex TE resonant frequencies in silicon
spheroid with permittivity $\epsilon=12$ for variation of aspect
ratio of polar $R_z$ and equatorial $R_{\bot}$ radii in the sector
$m=0$. Wave patterns show azimuthal component of electric field
$|E_{\phi}|$ of the Mie resonant modes in sphere at points marked
by closed circles where integers above the insets notify the
orbital momentum $l$ and the radial index $n$. 'x' marks the case
of oblate spheroid with $R_z=0.4R_{\bot}$ while '+' marks the case
of prolate spheroid with $R_z=1.6R_{\bot}$. } \label{fig1}
\end{figure}

The sector $m=0$ is simplified compared to the sector $m=1$
because of separation of TE and TM modes. Figure \ref{fig1}
presents evolution of complex TE resonant frequencies with
variation of the equatorial radius $R_{\bot}$ relative to the
polar radius $R_z$ from oblate silicon spheroid $R_z=0.4R_{\bot}$
to prolate spheroid $R_z=1.6R_{\bot}$. $k$ is the wave number, and
$R=(R_zR_{\bot}^2)^{1/3}$ is the mean radius that equalizes
volumes of sphere and spheroid. For the reader convenience we
split the frequency range in Figure \ref{fig1} into two parts. The
insets show the QNMs of a sphere.
\begin{figure}
\includegraphics*[width=8.5cm,clip=]{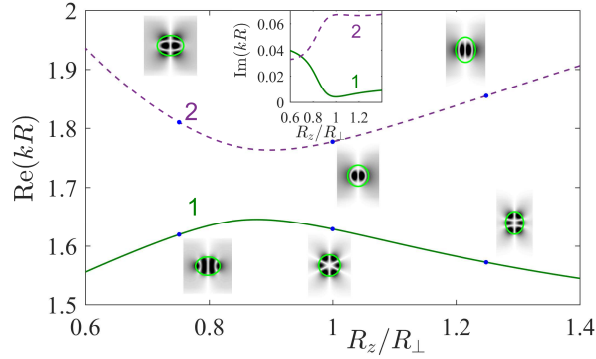}
\caption{Evolution of resonant frequencies and resonant modes
labelled as 1 and 2 in Fig. \ref{fig1} versus ratio of radii $R_z$
and $R_{\bot}$.} \label{fig2}
\end{figure}
In Figure \ref{fig2} we demonstrate a phenomenon of avoided
crossing of resonances marked as 1 and 2 in Figure \ref{fig1}
which is the result of interaction of the dipole QNM with the
octuple QNM
 \cite{Lai1991}. There is a general belief  that  a homogeneous
spherical  dielectric  body represents  the  ideal case,  so  that
any  perturbation of shape of sphere can only degrade  the
resonance (the imaginary part increases or the $Q$-factor
decreases). Lai {\it et al} \cite{Lai1990,Lai1991} have shown
this, however, provided that imaginary part of the spherical QNM
is small enough. For the QNMs with low $Q$-factor their
frequencies deviate from the complex eigenfrequencies of sphere
linearly \cite{Ching1998}.

This anomalous behavior of the low-$Q$ resonances  can be
comprehend if to refer to the series over spherical harmonics
(\ref{series}). For the TE polarization we have
\begin{equation}\label{TEseries}
    E_{\phi}=\sum_la_{l0}M_{l0}^{\phi}
\end{equation}
where $l=1, 3, 5, \ldots $  if $E_{\phi}$ is even relative to
$z\rightarrow -z$ and $l=2, 4, 6, \ldots $  if $E_{\phi}$ is odd.
Once a sphere transforms into spheroid the orbital momentum $l$ is
not preserved. Figure \ref{TEm0coef} shows as new multipole
radiation channels are opened with this transformation.
\begin{figure}
\includegraphics*[width=10cm,clip=]{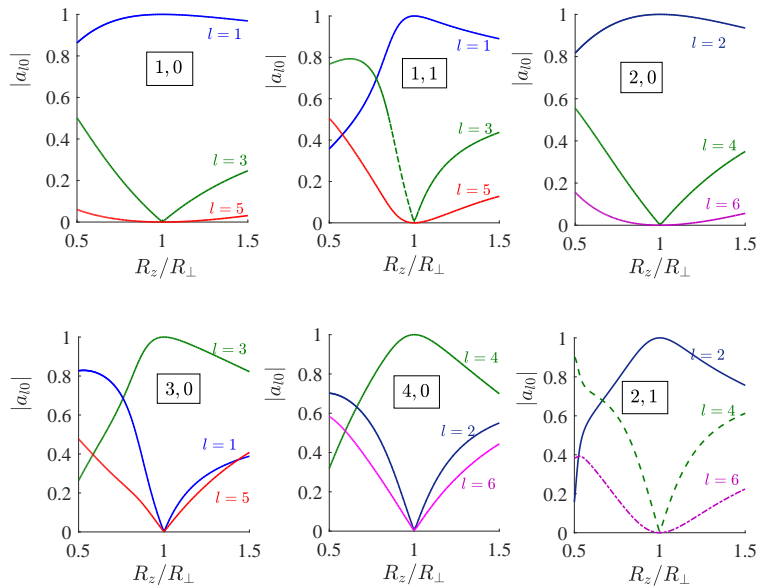}
\caption{Evolution of multipole coefficients in series
(\ref{series}) for evolution of resonant modes $l,n$ shown in Fig.
\ref{fig1}.} \label{fig3}
\end{figure}
Let us consider some of resonances shown Figure \ref{fig1}. For
variation of the polar radius $R_z$ the lowest mode shown by black
line goes through the Mie dipole mode $1,0$ of a sphere with the
frequency $kR=0.862+0.0414i$. As seen from the first subplot of
Figure \ref{fig3} at this moment the only radiation channel is
given by the coefficient $a_{10}$. The resonant widths of the Mie
resonant modes fast fall down with the orbital momentum $l$ and
grow with the radial index $n$ \cite{Lam1992}. As a result, when a
sphere is deformed, the fast decaying dipole channel is weakening
at the cost of linear arising of the next slower decay octuple
channel $l=3$ in accordance to Eq. (\ref{TEseries}). These
comprehensive considerations were issued by Lai {\it et al}
\cite{Lai1991}. Respectively the resonant width is decreased as
shown in Figure \ref{fig1} by black line. However, there are
exceptions from this rule, for example, the QNMs $l=2,n=1$ and
$l=2,n=0$ (The last column of subplots in Figure \ref{fig3}). In
both cases the same slower decaying radiation channels with $l=4$
and $l=6$ are attaching to the quadruple channel with $l=2$ for
deviation from a sphere. Nevertheless the behavior of resonant
widths is dramatically different as seen from Figure \ref{fig1}.
For the radial quantum $n=0$ we observe a degradation of the
quadruple QNM, while for $n=1$ we observe the opposite behavior.
That shows the importance of the radial indices for resonant
widths \cite{Lam1992}.

Let us consider also the resonances evolving with the Mie
resonances with higher orbital momentum, octuple resonance $3,0$
with the frequency $kR=1.629+0.0042i$ shown by  green line in
Figure \ref{fig1}. Corresponding evolution of multipole
coefficients is shown in Figure \ref{fig3} in subplot labelled
$3,0$. In contrast to previous dipole and quadruple resonances the
high-$Q$ decaying octuple resonance is substituted by the fast
decaying dipole resonance $1,0$. As a result we observe an
increase of resonant width in Figure \ref{fig1} for transformation
of sphere into spheroid. Other subplot $4,0$  in Figure \ref{fig3}
shows the same result.

\begin{figure}
\includegraphics*[width=8cm,clip=]{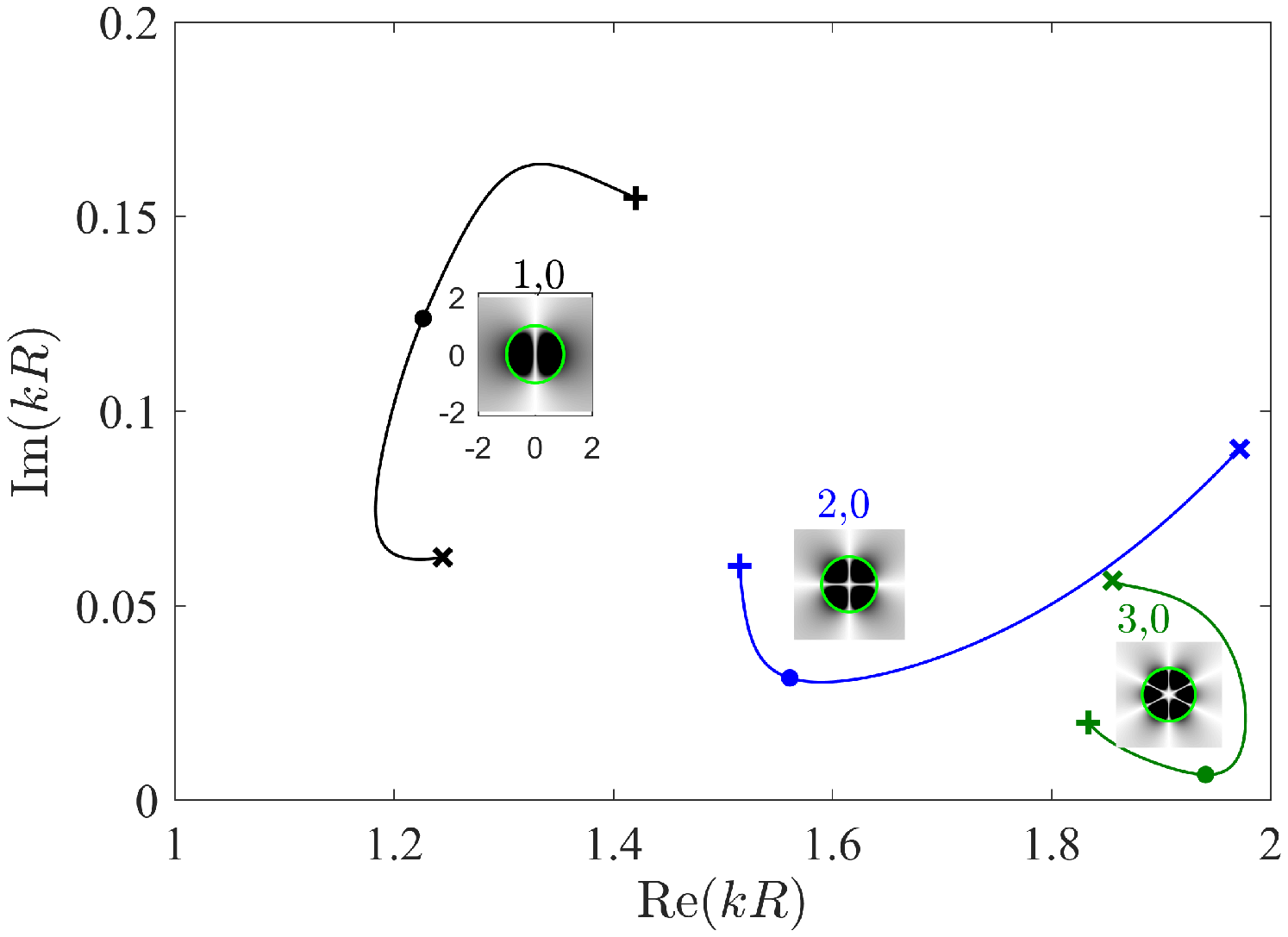}
\includegraphics*[width=8cm,clip=]{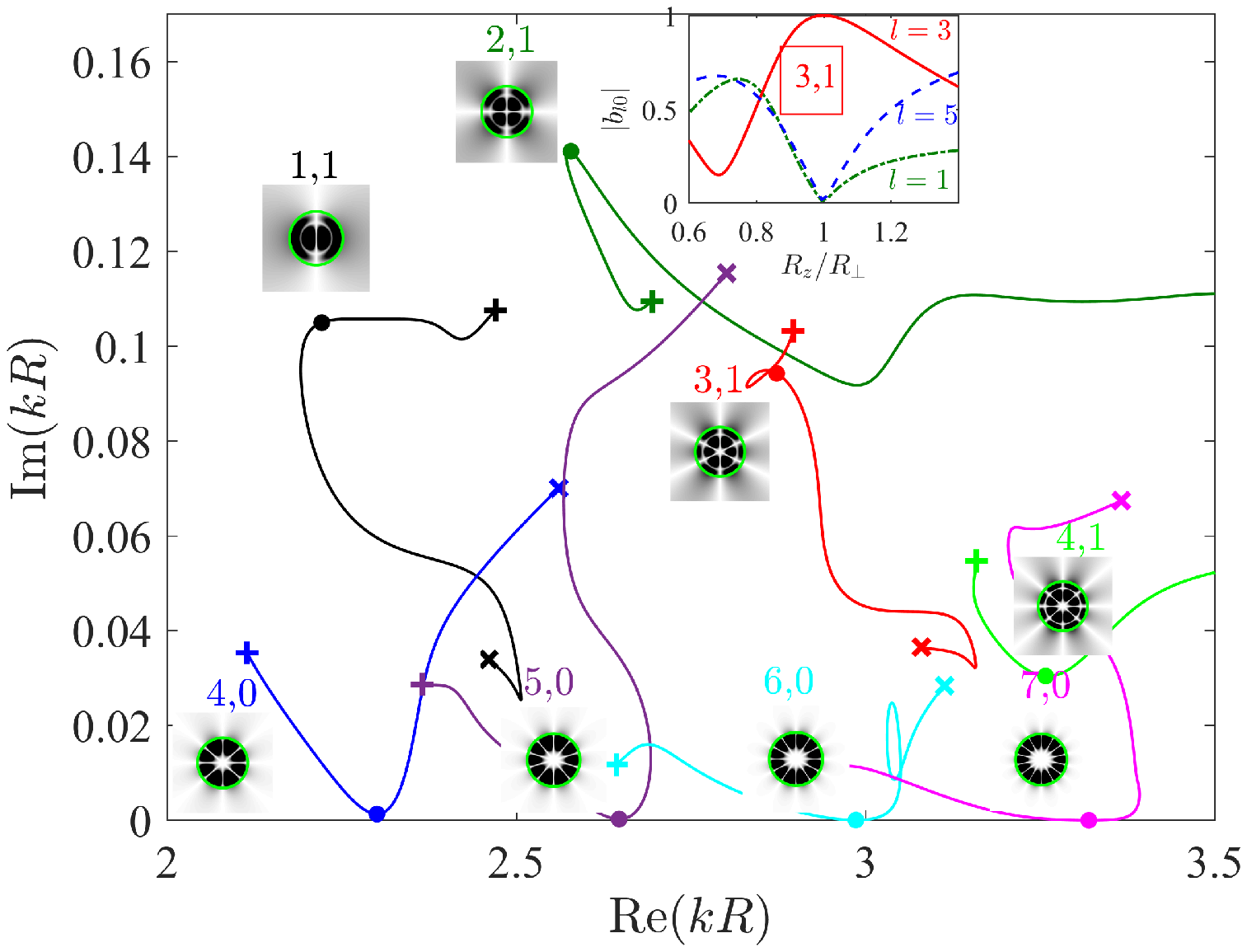}
\caption{Evolution of complex TM resonant frequencies. Wave
patterns show azimuthal component of magnetic field $|H_{\phi}|$
of the Mie resonant modes. 'x' marks $R_z=0.4R_{\bot}$ and '+'
marks $R_z=1.6R_{\bot}$. The inset shows behavior of multipolar
coefficients on the aspect ratio.} \label{fig4}
\end{figure}

We omit analysis of the TM resonances shown in Figure \ref{fig4}
because of a similarity with the case of the TE resonances except
that the series (\ref{series}) for magnetic field are given by the
coefficients $b_{l0}$ with the same sequence for $l=1, 3, 5,
\ldots$ for the even solutions of magnetic field $H_{\phi}$ and
$l=2, 4, 6, \ldots$ for the odd solutions relative to
$z\rightarrow -z$.
As a result we have similar rules for resonant widths. The Mie TM
dipole and quadrupole resonances yield to spheroid resonances in
the $Q$ factor in contrast to the Mie resonances with higher
orbital momenta. However there is an exception for  the resonance
$3,1$ which have no minimal resonant width at $R_z=R_{\bot}$. The
inset in Figure \ref{fig4} shows that in the prolate spheroid we
have extremely large contribution of the spherical harmonic $l=5$
compared to the dipole harmonic $l=1$ that suppresses emission
from the prolate spheroid.
\begin{figure}
    \includegraphics*[width=8cm,clip=]{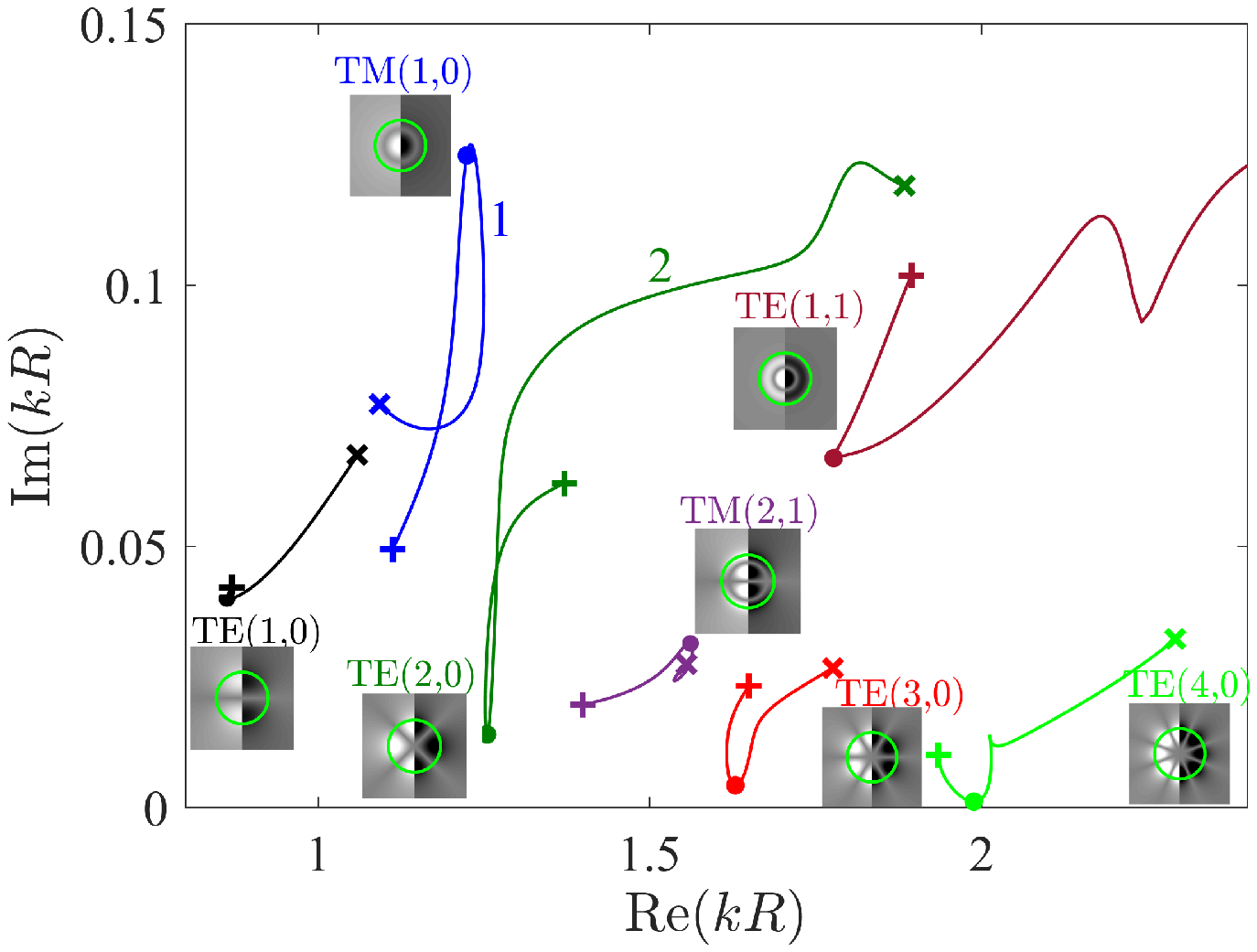}
    \includegraphics*[width=8cm,clip=]{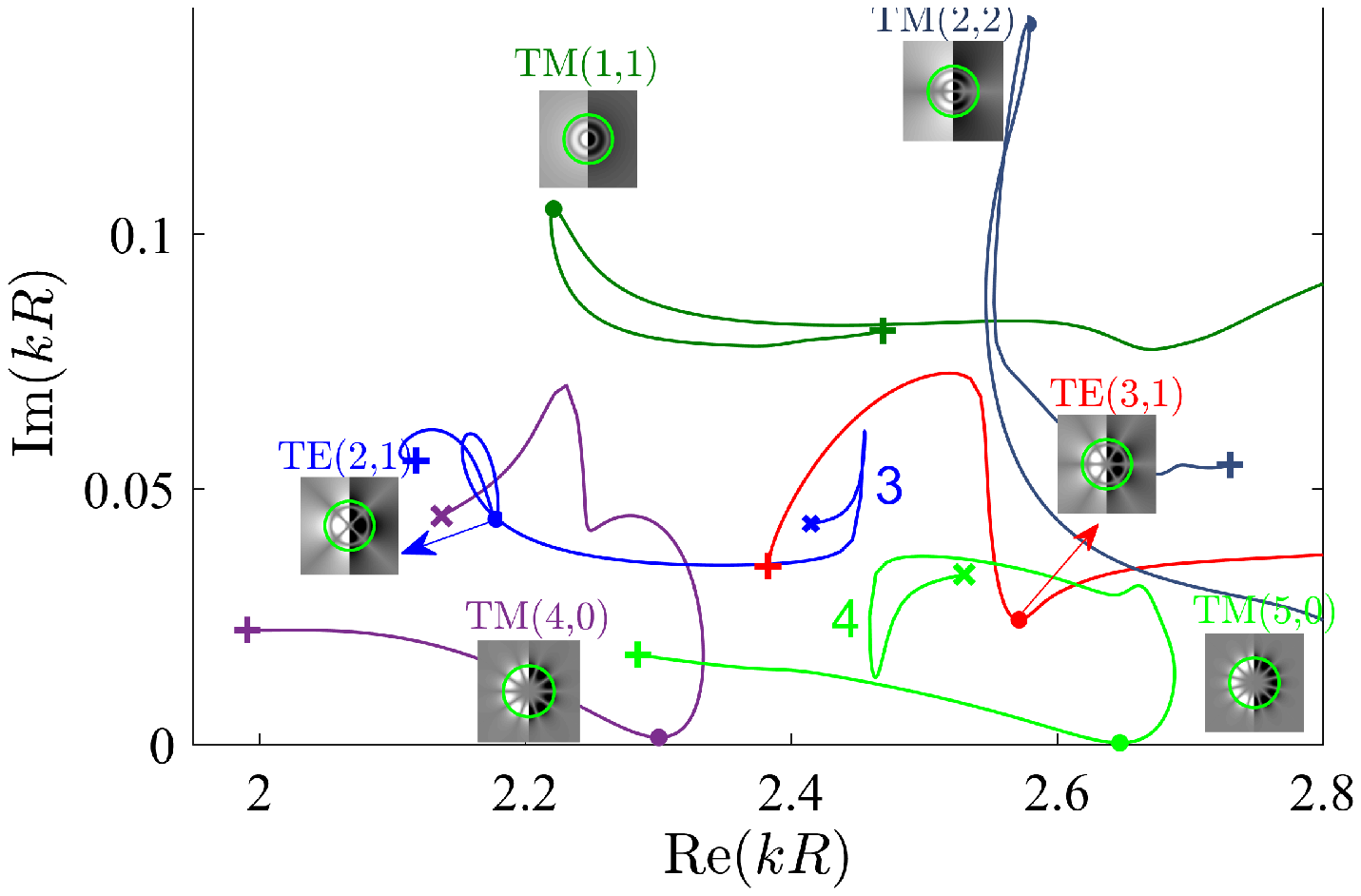}
    \caption{Evolution of resonant frequencies for traversing from the
        oblate spheroid $R_z=0.4R_{\bot}$ (pluses) through a sphere
        (closed circles) to the prolate spheroid $R_z=1.6R_{\bot}$
        (crosses) in the sector $m=1$. Titles above the insets indicate
        the orbital momentum $l$ and radial index $n$ (the number of radial nodal circles).
        The TE/TM modes are presented by the azimuthal components $|E_{\phi}|$/$|H_{\phi}|$.}
        \label{fig5}
\end{figure}
The sector $m=1$ is destined to show that the phenomena of ARCs
exist in the other sectors of the azimuthal index $m$, in
particular $m=1$ as demonstrated in Figure \ref{fig5}. Moreover
one can observe the same tendency of degradation of the high-$Q$
QNMs and, visa versa, enhancement of the $Q$-factor for the
low-$Q$ QNMs for deformation of sphere.

\section{Exceptional points.}
\begin{figure}
\includegraphics*[width=8cm,clip=]{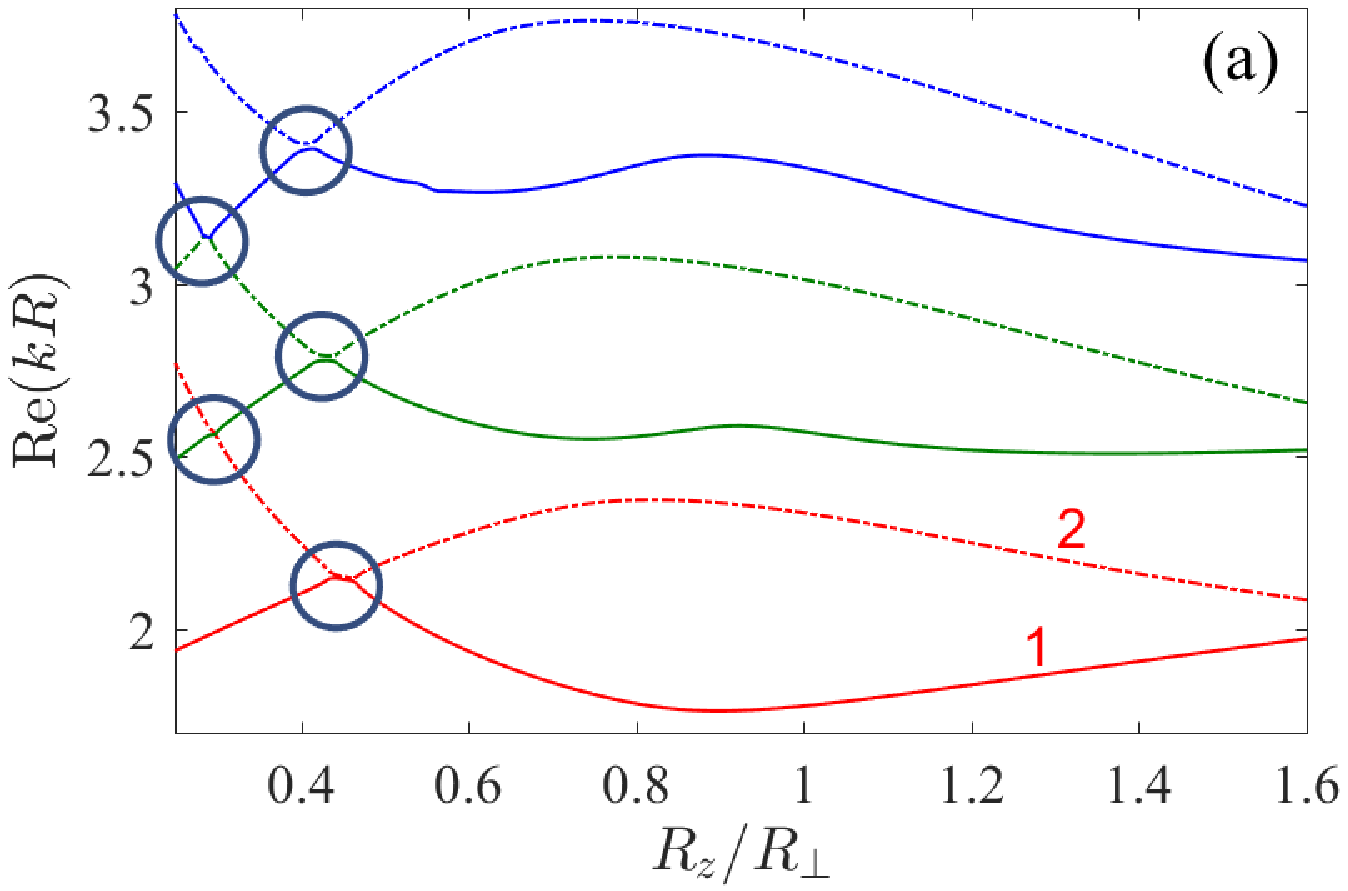}
\includegraphics*[width=7.1cm,clip=]{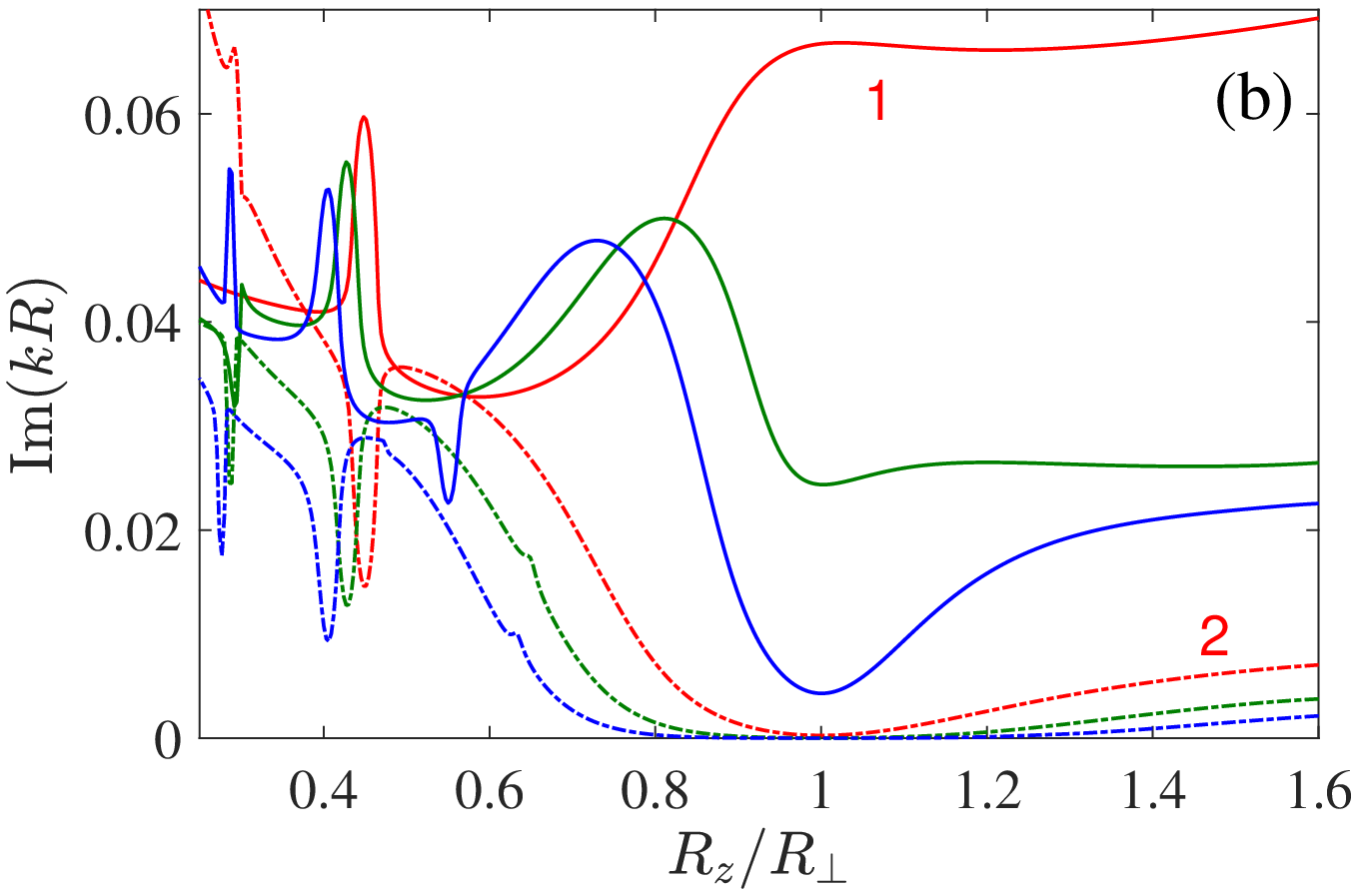}
\caption{ARCs of TE QNMs for evolution of sphere into spheroid in
the sector $m=0$.} \label{fig6}
\end{figure}

\begin{figure}
\includegraphics*[width=10cm,clip=]{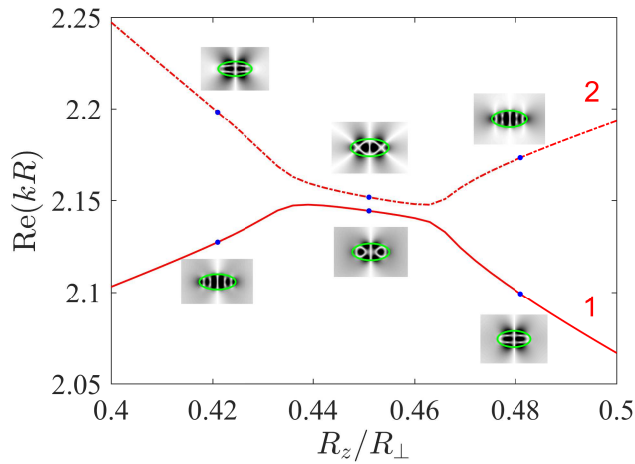}
\caption{Evolution of selected resonant frequencies and resonant
modes labelled  as 1 and 2 in Fig. \ref{fig6} vs ratio of radii
$R_z$ and $R_{\bot}$. The insets show the azimuthal component
$|E_{\phi}|$ of corresponding resonant modes at points marked by
closed circles.} \label{fig7}
\end{figure}

The sector $m=0$ demonstrates EPs separately for each
polarization. Figure \ref{fig6} shows numerous examples of avoided
crossing of TE modes highlighted by open circles.
It is interesting that the ARC phenomena are observed only for the
oblate spheroids below $R_z/R_{\bot}=1/2$.
The behavior of QNMs is presented in Figure \ref{fig7} which shows
as the  modes are exchanging for variation of the aspect ratio of
spheroid.
As shown in Figure \ref{fig6} (b) the ARCs are complemented by
strong enhancement of the $Q$-factor in an agreement with numerous
considerations in different dielectric resonators
\cite{Wiersig2006,Rybin2017,Chen2019a,Bulgakov2021}.

What is remarkable, the oblate spheroid demonstrates numerous  EPs
for the two-fold variation of the permittivity and the aspect
ratio for both sectors $m=0$ and $m=1$. Figure \ref{fig80} shows
the behavior of QNMs with the aspect ratio at $\epsilon=17.2$ in
the sector $m=0$. One can see that inside the areas highlighted by
open circles two QNMs coalesce into the one QNM.
\begin{figure}
\includegraphics*[width=10cm,clip=]{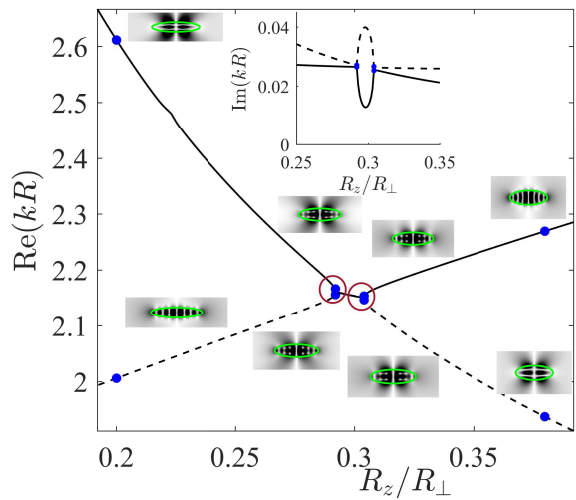}
\caption{Evolution of resonant frequencies and resonant modes
versus $R_z/R_{\bot}$ at $\epsilon=17.2$ in the sector $m=0$. Open
circles highlight EPs. The left one at $R_z/R_{\bot}=0.292,
\epsilon=17.2$ and the right one at $R_z/R_{\bot}=0.304,
\epsilon=18.4$. The insets show the $|E_{\phi}|$ profiles of TE
QNMs at points marked by closed circles.} \label{fig8}
\end{figure}
Such a behavior of resonances close to the EP behavior was
observed in different dielectric structures
\cite{Kullig2016,Ghosh2016,Yi2019,Huang2019a,Jiang2020}.
\begin{figure}
\includegraphics*[width=8cm,clip=]{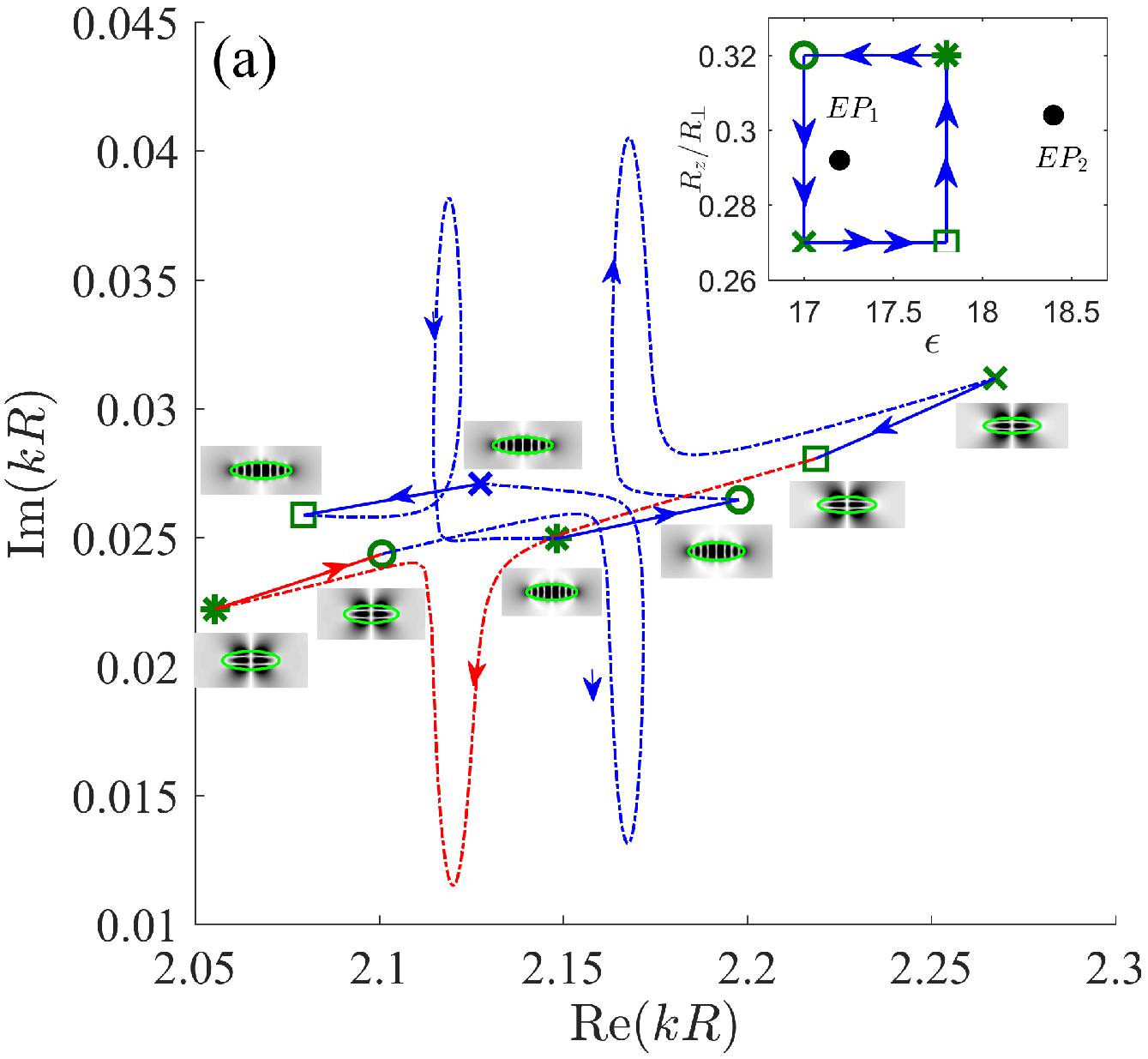}
\includegraphics*[width=8cm,clip=]{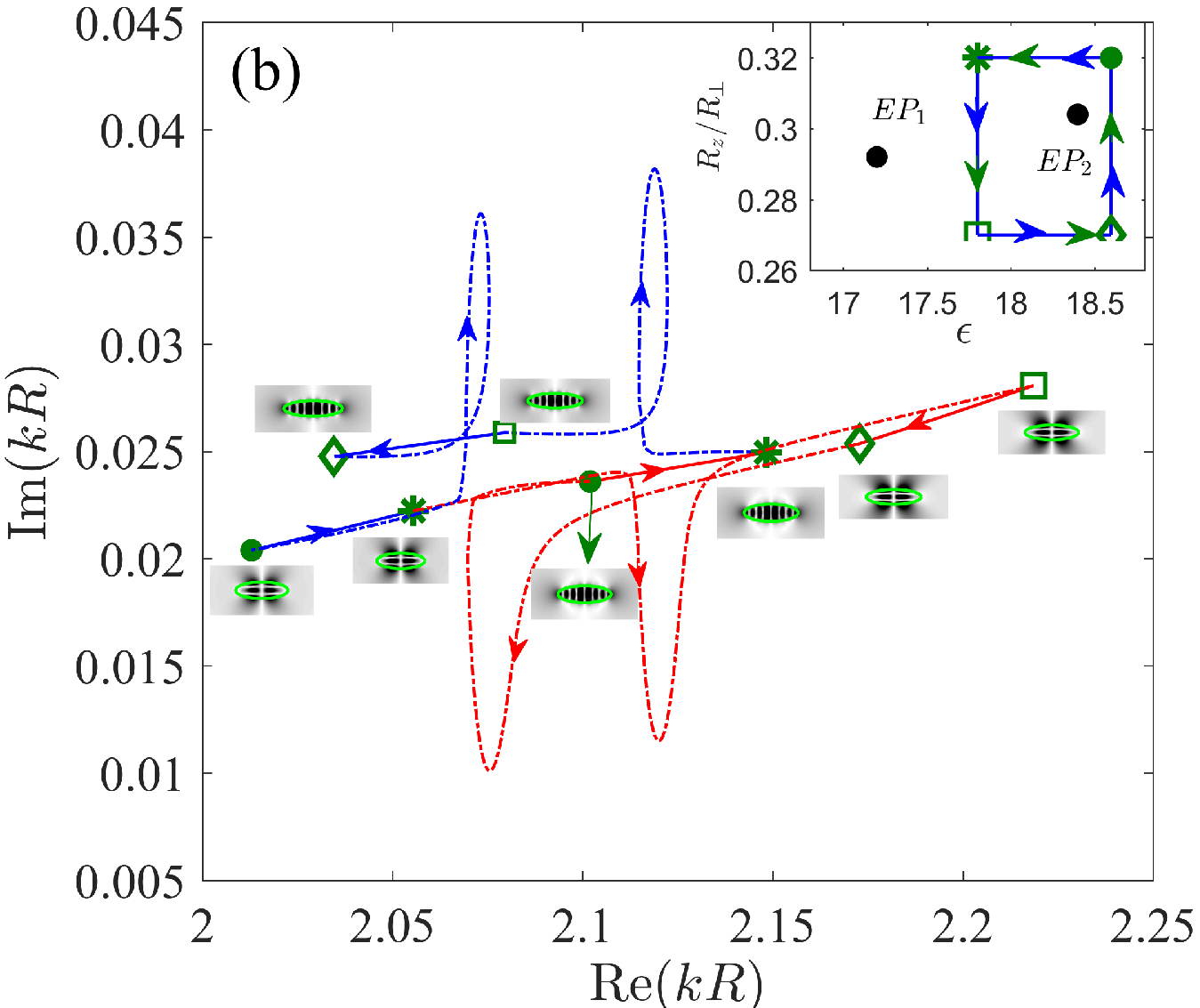}
\includegraphics*[width=8cm,clip=]{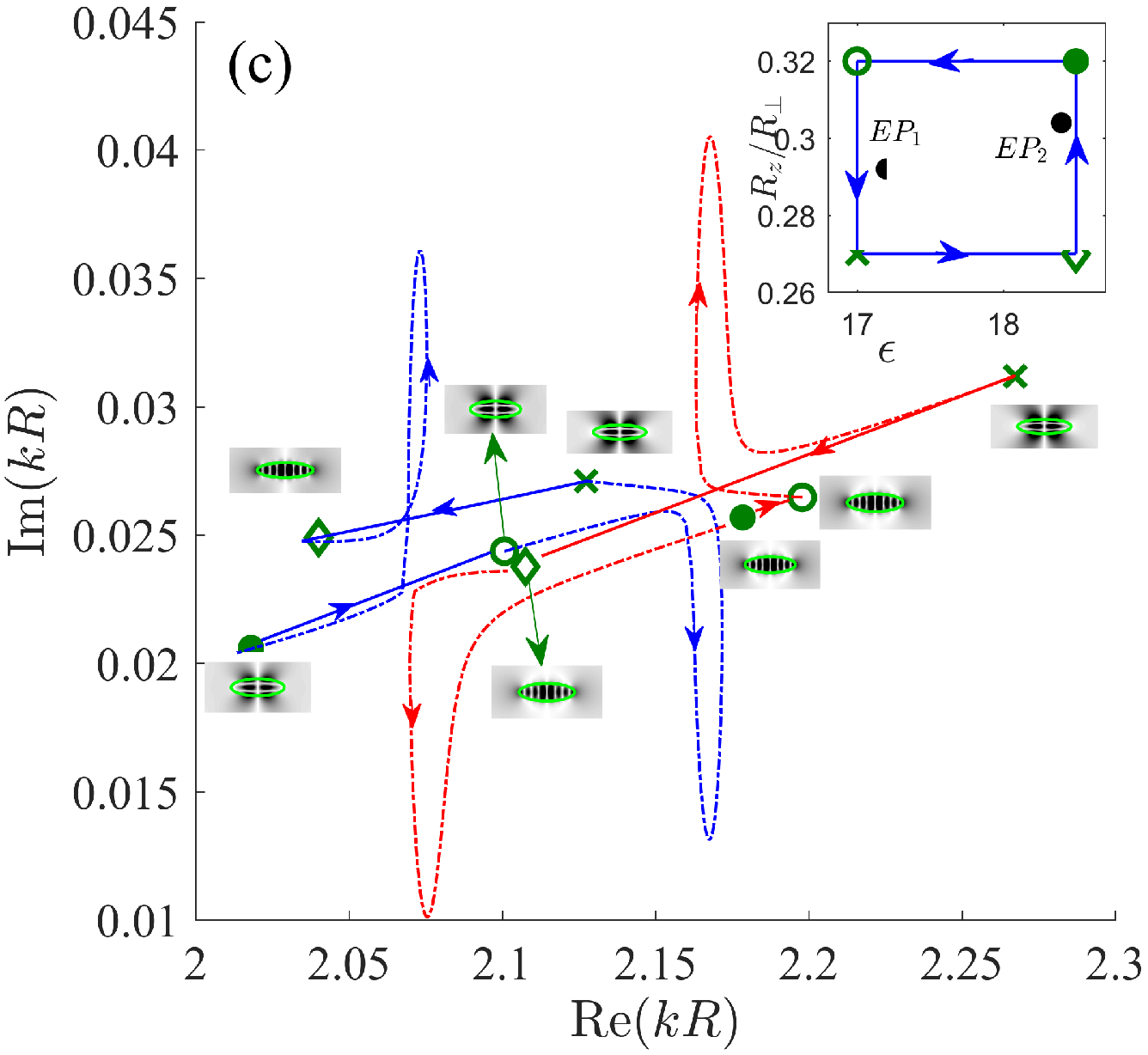}
\caption{Encircling of EPs shown by open circles in Fig.
\ref{fig80}. (a) and (b) Encircling separate EPs. (c) Encircling
of both EPs. Insets show the component $E_{\phi}$ of resonant
mode.} \label{fig9}
\end{figure}
In order to be convinced that there are indeed the EPs we encircle
the EP points shown open circles in Figure \ref{fig80} by three
ways. In the first case the rectangular contour encircles only the
left EP at the point $R_z/R_{\bot}=0.292, \epsilon=17.2$ as shown
in Figure \ref{fig9} (a). Respectively, in the second case the
contour encircles the right EP point $R_z/R_{\bot}=0.304,
\epsilon=18.4$ as shown in Figure \ref{fig9} (b). At last, we
present also the case of encircling of both EPs shown in Figure
\ref{fig9} (c). In all cases we encircle EPs counterclockwise.

Let us consider the first case shown in Figure \ref{fig9} (a)
where encircling starts with point $R_z/R_{\bot}=0.32,
\epsilon=17$ marked by open circle in the inset of Figure. In the
first downward path we decrease the aspect ratio  at the same
permittivity reaching the point till $R_z/R_{\bot}=0.27,
\epsilon=17$ marked by cross. In the complex plane this path maps
into sharp trajectory shown by dot-dashed blue line that features
high response of resonant frequency on shape of spheroid.
Respectively the resonant mode demonstrate sharp change of the
resonant mode. In the next horizontal path we slightly increase
the permittivity from $\epsilon=17$ till $\epsilon=17.8$ of the
oblate spheroid with the same shape and reach the point
$R_z/R_{\bot}=0.27, \epsilon=17.8$ marked by square in the inset.
In the complex plane this path maps into monotonic descent of
resonant frequency by law $(kR)^2\epsilon\approx C$ or $kR\approx
\sqrt{C/17}(1-\Delta\epsilon/2)$. That linear part of trajectory
is plotted by solid blue line in Figure \ref{fig9} (a). The
resonant mode presented by the insets at staring and finishing
points also does not show visible changes. The third upward part
of rectangular contour goes from the point marked by square
$R_z/R_{\bot}=0.27, \epsilon=17.8$ to the point marked by star
$R_z/R_{\bot}=0.32, \epsilon=17.8$ maps into sharp trajectory
shown by blue dash line. However the resonant mode is not changing
that is related to far distance between the left EP and the path
as distinct from the first downward path from circle to cross. By
doing so we closed the rectangular contour however as the resonant
frequency as the resonant mode are interchanged as was first
demonstrated by Dembowskii {\it et al} in a microwave metallic
resonator \cite{Dembowski2001}. And only the second encircling of
the left EP restores the resonant mode as demonstrated in Figure
\ref{fig9} (a) by red lines.

The right EP $R_z/R_{\bot}=0.304, \epsilon=18.4$ is expected to
give rise to the same features. However as shown in Figure
\ref{fig9} (b) counterclockwise encircling of this EP demonstrates
clockwise behavior of the resonant frequency and mode opposite to
the case of counterclockwise encircling of the left EP. That is
related to that the signs of winding numbers of neighboring EPs
arising after crossing of two lines in the complex plane are
opposite each other \cite{Shvartsman1994,Berggren2002}. Figure
\ref{fig9} (c) presents graphical evidence for that. The one
encircling of both EPs restores the resonant modes of each
resonance.

\begin{figure}
\includegraphics*[width=8cm,clip=]{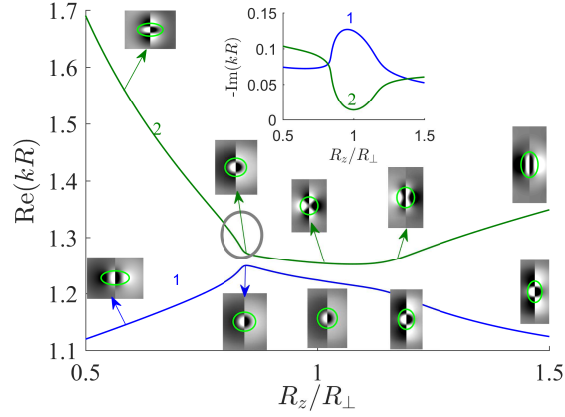}
\caption{Evolution of resonant frequencies and resonant modes
marked as 1 and 2  in Fig. \ref{fig5} (sector $m=1$) versus ratio
of radii $R_z$ and $R_{\bot}$ around EPs at $\epsilon=12$. The EP
is given by the point $\epsilon=12, R_z/R_{\bot}=0.84, kR=1.25$.}
\label{fig10}
\end{figure}

Next, we show an existence of EPs in the sector $m=1$ too in which
the QNMs with mixed polarizations can be excited by plane wave
incident along the z-axis as different from the case $m=0$
\cite{Stratton}. The first example of evolution of the QNMs (only
the component $E_{\phi}$ is presented) and their complex
eigenfrequencies in the sector $m=1$ is presented  in Figure
\ref{fig10}.

\begin{figure}
\includegraphics*[width=8cm,clip=]{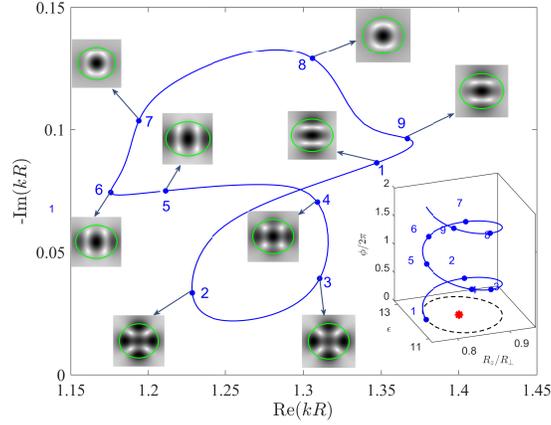}
\caption{Evolution of the field patterns $E_y$ for encircling the
EPs $\epsilon=12, R_z/R_{\bot}=0.84$ marked by star.}
\label{fig11}
\end{figure}

The EPs occur for precise two-fold tuning of the aspect ratio
$R_z/R_{\bot}$ and  the refractive index of spheroid that is
challengeable experimentally. However there is a way to show EPs
by encircling the EP through which resonant eigenmodes are
interchanged \cite{Dembowski2001}. Figure \ref{fig11} demonstrates
as for encircling of the EPs in plane $R_{\bot}/R_z$ and
$\epsilon$ one of resonant modes restores only after encircling by
$4\pi$. It is clear that the same refers to the multipole
coefficients $a_{l1}$ and $b_{l1}$ as shown in Figure \ref{fig3}.
\begin{figure}
\includegraphics*[width=8cm,clip=]{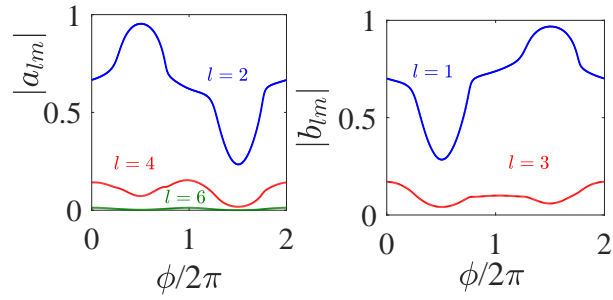}
\caption{Evolution of the expansion coefficients $a_{l1}$ (TE
modes) and $b_{l1}$ (TM modes) for encircling the EP shown in Fig.
\ref{fig11}.} \label{fig12}
\end{figure}

\begin{figure}
\includegraphics*[width=8cm,clip=]{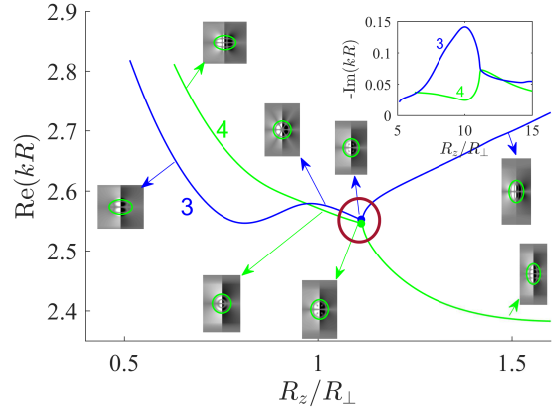}
\caption{Evolution of resonant frequencies and resonant modes
marked as 3 and 4  in Fig. \ref{fig5} (sector $m=1$) versus ratio
of radii $R_z$ and $R_{\bot}$ around EPs at $\epsilon=12$. The EP
is given by the point $\epsilon=12.46, R_z/R_{\bot}=1.11,
kR=2.56$.} \label{fig13}
\end{figure}

\begin{figure}
\includegraphics*[width=10cm,clip=]{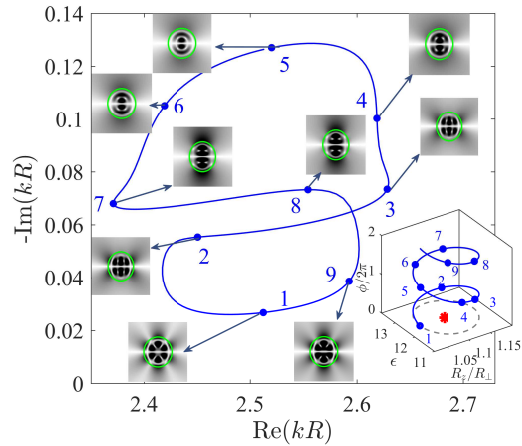}
\caption{Evolution of the field patterns $E_y$ for encircling the
EPs highlighted in Fig. \ref{fig13}.} \label{fig14}
\end{figure}

There are also many other EPs with higher frequencies. One example
of the EP is presented in Figures \ref{fig13} and \ref{fig14}.

\section{Summary and conclusions}
It seems reasonable that resonances of any dielectric particle
shaped differently from a sphere yield to the Mie resonances of
sphere by the $Q$-factors because the surface of sphere is
minimal. However as Lai {\it et al} \cite{Lai1990,Lai1991} have
shown that is truth only for those resonances whose imaginary part
is small enough. We present numerous examples which confirm this
rule and give comprehensible insight by demonstration of multipole
radiation channels for evolution of a sphere into spheroid.
However we also show exceptions from this rule.

 However the main objective of the present paper was demonstration of EPs in a
spheroid that has fundamental significance because of compactness
of these dielectric resonators. Moreover, evolution of expansion
coefficients in Fig\ref{fig6} demonstrate multipole conversion for
encircling of EPs and what is the most remarkable this evolution
has a period $4\pi$. In the photonic system, the appearance of EPs
can be exploited  to a broad range of interesting applications,
including lasing \cite{Feng2014}, asymmetric mode switching
\cite{Ghosh2016}, nonreciprocal light transmission
\cite{Feng2011,Laha2020}, enhancement of the spontaneous emission
\cite{Pick2017} and ultrasensitive sensing \cite{Chen2017}.

\begin{acknowledgments}
The work was supported by Russian Foundation for Basic Research
projects No. 19-02-00055.
\end{acknowledgments}
\bibliography{sadreev}
\end{document}